---------------------------------------------------------------------------
############################## .TEX FILE ###################################
---------------------------------------------------------------------------

\documentstyle[aps]{revtex}            
\begin{document}


\draft
\preprint{}
\title{Hypercomplex Group Theory}
\author{Stefano De Leo\thanks{{\em E-Mail}: 
              {\tt deleos@le.infn.it} {\footnotesize and} 
              {\tt deleo@ime.unicamp.br}    }}
\address{Dipartimento di Fisica, 
                Universit\`a degli Studi Lecce and INFN, 
                Sezione di Lecce\\
                via Arnesano, CP 193, 73100 Lecce, Italia\\
                and\\ 
         Instituto de Matem\'atica, Estat\'{\i}stica e 
                Computa\c{c}\~ao Cient\'{\i}fica, IMECC-UNICAMP\\
                CP 6065, 13081-970, Campinas, S.P., Brasil} 
\date{March, 1997}
\maketitle


\begin{abstract}

Due to the noncommutative nature of quaternions and octonions we introduce 
{\em barred operators}. This objects give the opportunity to manipulate 
appropriately the hypercomplex fields. The standard problems arising in the 
definitions of transpose, determinant and trace for quaternionic and
octonionic matrices are immediately overcome. We also investigate the 
possibility to formulate a {\em new approach} to Hypercomplex Group 
Theory (HGT). From a mathematical viewpoint, our aim is to highlight the
possibility of looking at new hypercomplex groups by the use of barred 
operators as fundamental step toward a clear and complete discussion of HGT.


\end{abstract}


\section{Introduction}
\label{s1}

Complex numbers have played a dual role in Physics, first as a technical tool
in resolving differential equation (e.g. in classical optics) or via the theory
of analytic functions for performing real integrations, summing series, etc.;
secondly in a more essential way in the development of Quantum Mechanics (and 
later Field Theory) characterized by complex wave functions and for fermions 
by complex wave equations. With quaternions, for the first type of 
application, i.e. as a means to simplify calculations, we can quote the 
original work of Hamilton~\cite{HAM}, 
but this only because of the late development of 
vector algebra by Gibbs and Heaviside~\cite{GH}. Even Maxwell used 
quaternions as a 
tool in his calculations, e.g. in the 
{\sl Treatise of Electricity and Magnetism}~\cite{MAX} 
we find the $\nabla$-operator expressed by the three quaternionic imaginary 
units.

Notwithstanding the 
Hamilton's conviction that quaternions would soon play a role 
comparable to, if not greater than, that of complex numbers the use of 
quaternions in Physics was very limited~\cite{GUR}. Nevertheless, in the 
last decades, we 
find a renewed interest in the application of noncommutative fields in 
Mathematics and Physics. In Physics, we quote quaternionic versions of
Gauge Theories~\cite{FIN1,HOR,QET,GUT}, 
Quantum Mechanics and Fields~\cite{FIN2,ADL1,ADL2,DIRAC,DKP,DR1},
Special Relativity~\cite{SR}. In Mathematics, we find applications of 
quaternions for Tensor Products~\cite{RH,DR2}, 
Group Representations~\cite{ADL3}. Nonassociative numbers are difficult to 
manipulate, nevertheless, the use
of the octonionic field within Quantum Mechanics~\cite{OQM}, in particular
in a formulation the Dirac Equation~\cite{ODE}, and in 
Group Theory~\cite{OREP} has recently appeared.

In this paper we aim to give a {\em new} panoramic review of hypercomplex 
groups. We use the adjective ``new'' since the elements of our matrices   
will not be simple 
quaternions or octonions but {\em barred hypercomplex operators}.

In Physics, particularly Quantum Mechanics, we are accustomed to 
distinguishing between ``states'' and ``operators''. Even when the operators
are represented by numerical matrices, the squared form of operators 
distinguishes them from the column structure of the spinors states. Only for 
one-component fields and operators is there potential confusion. In 
extending Quantum Mechanics defined over the complex field to quaternions
or even octonions, it has almost always been assumed that matrix operators
contain elements which are ``numbers'' indistinguishable from those of the 
state vectors. {\em This is an unjustified limitation}. In fact, 
(noncommutative) hypercomplex 
theories require barred operators~\cite{EVEN}.

This paper is organized as follows: In section~\ref{s2}, we introduce 
the quaternionic and octonionic algebras. In section~\ref{s3}, we show that 
the noncommutative nature of the quaternionic and octonionic fields suggest 
the use of  barred operators. We also give a brief review on the recent 
applications of barred operators in  
Mathematics and Physics. In section~\ref{s4} and~\ref{s5}, 
we find the appropriate 
definitions of transpose, trace and determinant for quaternionic and octonionic
matrices. Such sections also contain the {\em new} classification of
hypercomplex groups. Our conclusions are drawn in the final section.


\section{Hypercomplex Algebras}
\label{s2}

Complex numbers can be constructed from the real numbers by introducing
a quantity $e_1$ whose square is $-1$:
\[ c = r_1 + e_1 r_2 \quad \quad (r_{1,2} \in {\cal R}) \; .\]
Likewise, we can construct the quaternions from the complex numbers in exactly
the same way by introducing another quantity $e_2$ whose squared is $-1$,
\[ q = c_1 + e_2 c_2 \quad \quad (c_{1,2} \in {\cal C})\; ,\]
and which anticommutes with $e_1$ ($e_1e_2=-e_2e_1=e_3$). We wish to 
emphasize the need of {\em three 
anticommuting} imaginary units in constructing the quaternionic field 
(only two imaginary units are not sufficient to obtain Hamilton's 
field). 

In introducing the quaternionic algebra, let us follow the conceptual 
approach of Hamilton. In 1843, the Irish mathematician attempted to generalize
the complex field in order to describe the rotations in the three-dimensional 
space. He began by looking for numbers of the form 
\[x+e_1y+e_2z~,\] 
with $e_1^2=e_2^2=-1$. Hamilton's hope was to do for three-dimensional 
space what complex numbers do   
for the plane. Influenced by the existence of a complex number norm
\[ c^* c = (\mbox{Re} \, c)^2 +  (\mbox{Im} \, c)^2~,\]
when he looked at its generalization 
\[ (x-e_1y-e_2z) (x+e_1y+e_2z) = 
    x^2+y^2+z^2 - (e_1e_2+e_2e_1) yz~,\]
to obtain a real number, he had to adopt  the anticommutative law of 
multiplication for the imaginary units. Nevertheless, as remarked before, 
with only two imaginary 
units we have no chance of constructing a new 
numerical field, because assuming
\begin{eqnarray*}
e_1e_2=\alpha_0 + e_1 \alpha_1 + e_2 \alpha_2
~~~~~~~(\alpha_{0,1,2} \in {\cal R})~,\\
e_2e_1=\beta_0 + e_1 \beta_1 + e_2 \beta_2
~~~~~~~(\beta_{0,1,2} \in {\cal R})~,\\
\end{eqnarray*}
and
\[ e_1e_2=-e_2e_1~,\]
we find the relation
\[ \alpha_{0,1,2}= \beta_{0,1,2}=0~.\]
Thus, we must introduce a third imaginary unit $e_3\neq e_{1,2}$, with
\[ e_3=e_1e_2=-e_2e_1~. \] 
This noncommutative field is therefore characterized by three imaginary 
units $e_1$, $e_2$, $e_3$ which satisfy the following multiplication rules
\begin{equation}
e_1^2=e_2^2=e_3^2=e_1e_2e_3=-1~.
\end{equation}
Numbers of the form
\begin{equation}
\label{q}
q=x_0+ e_1x+e_2y+e_3z~~~~~~~~(~x_{0},x,y,z \in {\cal R}~)~,
\end{equation}
are called (real) {\em quaternions}. They 
are added, subtracted and multiplied according to the usual laws of 
arithmetic, except for the commutative law of multiplication.

Similarly to  rotations in a plane that can be concisely expressed by complex 
number, a
rotation about an axis passing through the origin and parallel to a given 
unitary vector $\hat{u}\equiv(u_x,u_y,u_z)$ by an angle $\alpha$ can be 
obtained taking the following {\em quaternionic} transformation
\[ \exp\left(\frac{\alpha}{2} \, \vec{e}\cdot \vec{u} \right)
\, \vec{e}\cdot \vec{r} \, 
\exp \left(-\frac{\alpha}{2} \, \vec{e}\cdot \vec{u} \right) ~ , \]
where $\vec{e}\equiv (e_1,e_2,e_3)$ and $\vec{r} \equiv (x,y,z)$. 
In section~\ref{s3}, 
we shall see how the quaternionic number $q$ in Eq.~(\ref{q}),
with the identification $x_0\equiv ct$, can be used to formulate a 
one-dimensional version of the Lorentz group~\cite{SR}. 
This gives the natural 
generalization of Hamilton's idea
\begin{center}
complex/plane~~~~~$\rightarrow$~~~~~
pure imaginary quaternions/space~~~~~$\rightarrow$~~~~~
quaternions/space-time~,
\end{center}
completing the unification of algebra and geometry.

Let us now consider the conjugate of $q$ 
\begin{equation}
\label{qb} 
q^{\dag} = x_0 - e_1 x - e_2 y - e_3 z~.
\end{equation}
We observe that $q^{\dag}q$ and $qq^{\dag}$ are both equal to the real number
\[ N(q) = x_0^2 + x^2 + y^2 + z^2~,\]
which is called the norm of $q$. When $q\neq 0$, we can define
\[ q^{-1} = q^{\dag} / N(q)~,\]
so the quaternions form a zero-division ring. Such a 
noncommutative number field is denoted, in Hamilton honour, by $\cal H$.

An important difference between quaternionic and complex numbers is related 
to the definition of the conjugation operation. Whereas with complex 
numbers we can define only one type of conjugation
\[ e_1 \rightarrow -e_1~,\]
working with quaternionic numbers we can introduce 
different conjugation operations.  
Indeed, with three imaginary units we
have the possibility to define besides the standard conjugation~(\ref{qb}), 
 the six new operations
\begin{eqnarray*}
(e_1,e_2,e_3) & ~\rightarrow ~& 
(-e_1,+e_2,+e_3)~,~(+e_1,-e_2,+e_3)~,~(+e_1,+e_2,-e_3)~;\\
(e_1,e_2,e_3) & ~\rightarrow ~& 
(+e_1,-e_2,-e_3)~,~(-e_1,+e_2,-e_3)~,~(-e_1,-e_2,+e_3)~.
\end{eqnarray*}
These last six conjugations can be concisely represented by $q$ and 
$q^{\dag}$ as follows
\begin{eqnarray*}
q & ~\rightarrow ~& -e_1q^{\dag}e_1~,~-e_2q^{\dag}e_2~,~-e_3q^{\dag}e_3~,\\
q & ~\rightarrow ~& -e_1qe_1~~,~-e_2qe_2~~,~-e_3qe_3~~.
\end{eqnarray*}
It could seem that the only independent conjugation be represented by 
$q^{\dag}$. Nevertheless, $q^{\dag}$ can also be expressed 
in terms of  $q$, in fact 
\begin{equation}
\label{qbar}
q^{\dag} = -\frac{1}{2} \; (q+e_1qe_1+e_2qe_2+e_3qe_3)~.
\end{equation}

In going from the complex numbers to the quaternions we lose the property of 
commutativity (e.g. $e_1e_2=-e_2e_1$). In going from the quaternions to 
the next more complicated division algebra, we also lose the property of 
associativity. Octonionic numbers can be constructed from the quaternions 
by introducing a new imaginary unit $e_4$ which anticommutes with the 
quaternionic imaginary units $e_1$, $e_2$ and $e_3$,
\[o=q_1+e_4 q_2 ~~~~~~~ (q_{1,2} \in {\cal H})~. \]
We can immediately show the nonassociativity of the octonionic numbers in the 
previous ``split'' representation. In fact, by starting from the 
seven imaginary units
\[ e_1,~e_2,~e_3,~e_4~,e_5=e_1e_4,~e_6=e_2e_4,~e_7=e_3e_4~,\]
it is straightforward to verify that 
\[
e_5e_6e_3=e_6e_7e_1=e_7e_5e_2=-1 \; , \]
in fact imposing associativity we have for example
\[ e_5e_6e_3=e_1e_4e_2e_4e_3=-e_1e_2e_4^2e_3=e_1e_2e_3=-1~.\]
Associativity fails in the following relations
\[ e_1(e_4e_3)=-e_1e_7=e_6~~~~~ \mbox{and}~~~~~
   (e_1e_4)e_3=e_5e_3=-e_6~.\]

We now summarize our notation for the octonionic algebra and 
introduce useful elementary properties to manipulate the nonassociative 
numbers. An octonionic number will be represented by
\begin{equation}
o = r_{0}+\sum_{m=1}^{7} r_{m}e_{m} \quad \quad (~r_{0,...,7}~~ 
\mbox{reals}~) 
\; , 
\end{equation}
where $e_{m}$ are elements obeying the noncommutative and nonassociative 
algebra
\begin{equation}
e_{m}e_{n}=-\delta_{mn}+ \epsilon_{mnp}e_{p} \quad \quad 
(~\mbox{{\footnotesize $m, \; n, \; p =1,..., 7$}}~) 
\; ,
\end{equation}
with $\epsilon_{mnp}$ totally antisymmetric and equal to unity for the seven 
combinations 
\[
123, \; 145, \; 176, \; 246, \; 257, \; 347 \; \mbox{and} \; 365  
\]
(each cycle represents a 
quaternionic subalgebra). Amongst the different (equivalent) possibilities, we
choose the previous combinations to uniform the notation of this paper with the
notation which appears in recent works~\cite{OQM,ODE,OREP}. 

The norm, $N(o)$, for the octonions is 
defined by
\begin{equation}
N(o)=(o^{\dag}o)^{\frac{1}{2}}=
(oo^{\dag})^{\frac{1}{2}}=
(r_{0}^{2}+  ... + r_{7}^{2})^{\frac{1}{2}} \; ,
\end{equation}
with the octonionic conjugate $o^{\dag}$ given by
\begin{equation}
o^{\dag}=r_{0}-\sum_{m=1}^{7} r_{m}e_{m} \; . 
\end{equation} 
The inverse is then 
\begin{equation}
o^{-1}=o^{\dag}/N(o) \quad \quad (~o\neq 0~)~.
\end{equation}
We can define an {\em associator} (analogous to the usual algebraic 
commutator) as follows
\begin{equation}
\label{ass}
\{x, \; y , \; z\}\equiv (xy)z-x(yz) \; ,
\end{equation}
where, in each term on the right-hand, we must, first of all, perform 
the multiplication in brackets. 
Note that for real, complex and quaternionic numbers the associator is 
trivially null. For octonionic imaginary units we have
\begin{equation}
\label{eqass}
\{e_{m}, \; e_{n}, \; e_{p} \}\equiv(e_{m}e_{n})e_{p}-e_{m}(e_{n}e_{p})=
2 \epsilon_{mnps} e_{s} \; ,
\end{equation}
with $\epsilon_{mnps}$ totally antisymmetric and equal to unity for the seven 
combinations 
\[ 
1247, \; 1265, \; 2345, \; 2376, \; 3146, \; 3157 \; \mbox{and} \; 4576 \; .
\]
Working with octonionic numbers the associator~(\ref{ass}) is in general 
non-vanishing, however, an ``alternative condition'' is fulfilled
\begin{equation}
\label{rul}
\{ x, \; y, \; z\}+\{ z, \; y, \; x\}=0 \; .
\end{equation}


\section{Barred Operators}
\label{s3}

Due to the noncommutative nature of quaternions we must distinguish between 
$q_1q_2$ and $q_2q_1$. Thus,  it is appropriate to consider left/right-actions
for our imaginary units $e_1$, $e_2$ and $e_3$. We 
introduce {\em barred operators}~\cite{EVEN} to represent, in a compact way, 
the right-action of the three quaternionic imaginary units. Explicitly, we 
write
\begin{equation}
1\mid e_1 \; , \; 1\mid e_2 \; , \; 1\mid e_3
\end{equation}
to identify the right multiplication of $e_1$, $e_2$, $e_3$ and so
\[ (1\mid e_m) q \equiv q e_m~~~~~~~
(\mbox{{\footnotesize $m=1,2,3$}})~.\]
In this formalism, the most general transformation on quaternions 
will be given by
\begin{equation}
\label{barred}
q_0 + q_1 \mid e_1 + q_2 \mid e_2 + q_3 \mid e_3
\quad \quad (q_{0,1,2,3} \in {\cal H}) \; .
\end{equation}

In the last few years the left/right-action of the quaternionic numbers, 
expressed by barred operators~(\ref{barred}), has been very
useful in overcoming difficulties owing to the noncommutativity of
quaternions. Among the successful applications of barred
operators we mention the one-dimensional quaternionic formulation of
Lorentz boosts. Explicitly, the quaternionic generators of the Lorentz 
group are
\begin{center}
\begin{tabular}{lc}
boost $(ct,x)$ & $\frac{e_3 \mid e_2 - e_2 \mid e_3}{2}$~,\\
boost $(ct,y)$ & $\frac{e_1 \mid e_3 - e_3 \mid e_1}{2}$~,\\
boost $(ct,z)$ & $\frac{e_2 \mid e_1 - e_1 \mid e_2}{2}$~,\\
rotation around $x$~~~~~~~ & $\frac{e_1 - 1 \mid e_1}{2}$~,\\
rotation around $y$        & $\frac{e_2 - 1 \mid e_2}{2}$~,\\
rotation around $z$        & $\frac{e_3 - 1 \mid e_3}{2}$~.
\end{tabular}
\end{center}
The four real quantities which identify the space-time point $(ct,x,y,z)$
are represented by the quaternion
\[ q=ct + e_1 x + e_2 y + e_3 z~.\]
We know that in analogy to the connection between the rotation group
$O(3)$ and the special unitary group $SU(2)$, there is a natural correspondence
between the Lorentz group $O(3,1)$ and the special linear group $SL(2)$. The 
use of barred operators~(\ref{barred}) gives us the possibility to extend the 
connection between the special unitary group $SU(2)$ and the unitary 
quaternions by allowing 
a {\em one-dimensional} quaternionic version of the special linear group
$SL(2)$ (a detailed discussion is found in ref.~\cite{SR}). We also note that
barred operators~(\ref{barred}) have $16$ real parameters, the same number of
parameters which appear in $4\times 4$ real matrices. This suggests a 
correspondence between such barred quaternionic operators and generic
$4$-dimensional real matrices (appendix~A). 

New possibilities, coming out from the use of barred operators, also appear
in Quantum Mechanics and Field Theory, e.g. they allow an appropriate
definition of the momentum operator~\cite{DIRAC}, quaternionic version of 
standard relativistic equations~\cite{DIRAC,DKP}, Lagrangian 
formalism~\cite{LAG}, electroweak model~\cite{QET} and grand unification 
theories~\cite{GUT}.

Let us now discuss the algebra of barred operators and introduce some 
elementary relations and definitions which will be useful in the following 
sections. Remembering the noncommutativity of the quaternionic 
multiplication, we must 
specify if our scalar factors are quaternionic, complex or real numbers.
Operators which act on states {\em only} from the left (i.e. quaternionic 
numbers) will be named {\em quaternionic linear operator} and will be simply 
indicated by $q$. Obviously, from these more general classes of operators, 
such as complex or real linear quaternionic operators, can be constructed. 
For example, the barred operator~(\ref{barred}) represents a {\em real linear
quaternionic operator}. It will be denoted by ${\cal Q}_r$,
\[
{\cal Q}_r = q_0 + q_1 \mid e_1 + q_2 \mid e_2 + q_3 \mid e_3 \; .
\]
To complete the list of 
possible barred operators we give an explicit example of 
{\em complex linear quaternionic operator}
\[ {\cal Q}_c \equiv q_0 + q_1 \mid e_1~.\]

In this section, we shall deal with the algebra of real linear quaternionic 
operators
\[ {\cal Q}_r \supset {\cal Q}_c \supset q~. \]
Our considerations and conclusions can be immediately translated
to complex linear quaternionic operator. 

The product of two barred operators 
${\cal Q}_r$ and ${\cal P}_r$ in terms of quaternions $q_{0,1,2,3}$ and 
$p_{0,1,2,3}$ is given by
\begin{eqnarray*} 
{\cal Q}_r {\cal P}_r & = & 
q_0p_0 - q_1p_1 - q_2 p_2 - q_3p_3 +\\
 &  & (q_0p_1 + q_1p_0 - q_2 p_3 + q_3p_2) \mid e_1 +\\
 &  & (q_0p_2 + q_2p_0 - q_3 p_1 + q_1p_3) \mid e_2 +\\
 &  & (q_0p_3 + q_3p_0 - q_1 p_2 + q_2p_1) \mid e_3 ~.
\end{eqnarray*}
The ``full'' conjugation operation is defined by changing the sign of our
left/right quaternionic imaginary units, i.e.
\[ (e_1,e_2,e_3)^{\dag}=-(e_1,e_2,e_3) ~~~~~~~~
\mbox{and} ~~~~~~~
(1\mid e_1,1\mid e_2,1\mid e_3)^{\dag}=-(1\mid e_1,1\mid e_2,1\mid e_3)~.
\]
The previous definition implies
\begin{eqnarray*} 
[(q_1\mid q_2)(p_1 \mid p_2)]^{\dag} & = & (q_1p_1\mid p_2q_2)^{\dag} =
p_1^{\dag} q_1^{\dag} \mid q_2^{\dag} p_2^{\dag} \\
 & = & (p_1 \mid p_2)^{\dag} (q_1\mid q_2)^{\dag}~,
\end{eqnarray*}
and so 
\[ \left( {\cal Q}_r {\cal P}_r \right)^{\dag} = 
{\cal P}_r^{\dag}{\cal Q}_r^{\dag}~.
\]

In section~\ref{s4}, dealing with quaternionic matrices 
we shall distinguish between real linear
quaternionic groups, $GL(n,{\cal Q}_r)$, and complex linear quaternionic 
groups, $GL(n,{\cal Q}_c)$. For a clear and complete discussion of standard 
quaternionic groups, $GL(n,q)$, the reader is referred to 
Gilmore's book~\cite{GIL}. The use of
barred operators give new opportunities in HGT. Let us observe as follows.
The so-called ``symplectic'' complex representation of a quaternion (state) 
$q$
\[ q=c_1 + e_2 c_2~~~~~~~(c_{1,2} \in {\cal C})~,\]  
by a complex column matrix, is
\begin{equation}
\label{sym}
q ~\leftrightarrow ~
\left( \begin{array}{c} c_1\\ c_2 \end{array} \right)~. 
\end{equation}
The operator representation of $e_1$, $e_2$ and $e_3$ consistent with the 
above identification
\begin{equation}
\label{ident}
   e_1 \leftrightarrow \left( \begin{array}{cc} i & 0\\ 0 & -i
                              \end{array} \right)=i\sigma_3~,~~~~~
   e_2 \leftrightarrow \left( \begin{array}{cc} 0 & -1\\ 1 & 0
                              \end{array} \right)=-i\sigma_2~,~~~~~
   e_3 \leftrightarrow \left( \begin{array}{cc} 0 & -i\\ -i & 0
                              \end{array} \right)=-i\sigma_1~,
\end{equation}
has been known since the discovery of quaternions. It permits any 
quaternionic number or matrix to be translated into a complex matrix, 
{\em but not necessarily viceversa}. Eight real numbers are required to define
the most general $2\times 2$ complex matrix but only four are needed to define
the most general quaternion. In fact since every (non-zero) quaternion has an 
inverse, only a subclass of invertible $2\times 2$ complex matrices are
identifiable with quaternions. Complex linear quaternionic operators complete 
the translation~\cite{EVEN}. 
The barred quaternionic imaginary unit
\[ 1\mid e_1 ~\leftrightarrow ~ \left( \begin{array}{cc} i & 0\\ 0 & i
                                       \end{array} \right)~,\]
adds four additional degrees of freedom, obtained by matrix multiplication
of the corresponding matrices,
\[ 1\mid e_1~,~e_1 \mid e_1~,~e_2 \mid e_1~,~e_3 \mid e_1~,\] 
and so we have a set of rules for translating from any $2\times 2$ complex 
matrices to ${\cal Q}_c$-barred operators. This opens new possibilities
for quaternionic numbers, see for example 
the one-dimensional version of the 
Glashow group~\cite{QET}. Obviously this translation does not apply to 
odd-dimensional complex matrices~\cite{ODD}.

We conclude this section by extending our considerations on barred operators
to the octonionic field. Here, the situation is more delicate. Due to the 
nonassociativity of octonions we must distinguish between {\em left} and 
{\em right} barred operator. The natural octonionic extension 
of~(\ref{barred}) should be 
\begin{equation}
o_{0}+ \sum_{m=1}^{7} o_{m}\mid e_{m} ~~~~~~~
(o_{0, ...,7}~~ \mbox{octonions})~,
\end{equation}
but, due to the nonassociativity, this operator is not a well defined object.
For example, the triple product $o_1oe_1$ could be either 
$(o_1o)e_1$ or $o_1(oe_1)$. So, in order to avoid ambiguities, 
we need to define {\em left/right-barred 
operators}. Left barred operators will be indicated by
\[ o_1~)~e_1 + o_2~)~e_2 + ... + o_7~)~e_7 ~,\]
in similar way we introduce right barred operators
\[ o_1~(~e_1 + o_2~(~e_2 + ... + o_7~(~e_7 ~.\]
Their action on a generic quaternionic number $o$ is respectively represented 
by
\[ (o_1o)e_1 + (o_2o)e_2 + ... + (o_7o)e_7 ~,\]
and
\[ o_1(oe_1) + o_2(oe_2) + ... + o_7(oe_7) ~.\]
Nevertheless, there are barred operators in which the nonassociativity does 
not apply, like
\[ 1~)~e_m = 1~(~e_m \equiv 1\mid e_m ~ , \]
or
\[ e_m~)~e_m = e_m~(~e_m \equiv e_m\mid e_m ~ . \]
The counting of independent barred operators should be 106, explicitly
\begin{center}
\begin{tabular}{lr}
$1, ~ e_{m},~  1\mid e_{m}$ & {\footnotesize ~~~~~(15 elements)} ,\\
$e_{m}\mid e_{m}$ &{\footnotesize (7)} ,\\
$e_{m}~)~e_{n}$~~~~{\footnotesize $(m\neq n)$} & {\footnotesize (42)} ,\\
$e_{m}~(~e_{n}$~~~~{\footnotesize $(m\neq n)$} & {\footnotesize (42)} ,\\
{\footnotesize $m=1,2,...,7$} &
\end{tabular}
\end{center}
Yet, we can prove that each right-barred operator can be 
expressed by a suitable combination of left-barred operators. The proof is 
based on the correspondence between the left-right barred octonionic 
operators and generic $8\times 8$ real matrices~\cite{OREP}. 
In appendix~B, we report, 
for the sake of completeness, the translation rules between octonionic 
left-right barred operators 
and $8 \times 8$ real matrices. 
Thus, to represent the most general 
octonionic operator, we need only left-barred objects 
\begin{equation}
\label{obarred}
o_{0}+\sum_{m=1}^{7} o_{m}~)~e_{m} ~,
\end{equation}
reducing to 64 the previous 106 elements. Barred operators~(\ref{obarred}),
which will be denoted by ${\cal O}_r$, represent {\em real linear octonionic
operators}. From these, we can immediately extract 
{\em complex linear octonionic operator}, ${\cal O}_c$, and obviously the
standard {\em octonionic linear operators}, $o$,
\[ {\cal O}_r \supset {\cal O}_c \supset o~. \]

The classification of octonionic/quaternionic operator, given in this section,
can be concisely summarized as follows
\begin{center}
\begin{tabular}{|l||c|c||c|}
\hline
~~Number Field~~ & ~~Real Linear B.O.~~   & ~~Complex Linear B.O.~~ & 
~~State~~\\ \hline
~~Octonionic   & ${\cal O}_r$ {\footnotesize (64)}  & 
${\cal O}_c$ {\footnotesize (16)}   & 
$o$ {\footnotesize (8)}\\ \hline
~~Quaternionic & 
${\cal Q}_r$ {\footnotesize (16)}  & 
${\cal Q}_c$ {\footnotesize (8)}   & 
$q$ {\footnotesize (4)}\\ \hline
\end{tabular}
\\ \vspace*{.3cm}
{\footnotesize B.O. $\leftrightarrow$ Barred Operator,}
\end{center}
in parenthesis we recall the number of real parameters characterizing
the respective barred operator.


\section{Quaternionic Groups}
\label{s4}

Every set of basis vectors in $V_n$ can be related to every other coordinate
system by an $n\times n$ non singular matrix. The $n\times n$ matrix groups 
involved in changing bases in the vector spaces ${\cal R}_n$, ${\cal C}_n$ and
${\cal H}_n$ are called {\em general linear groups} of $n\times n$ matrices
over the reals, complex and quaternions
\begin{center}
\begin{tabular}{ccccc}
$GL(n,r)$ & ~~~~~$\rightarrow$~~~~~ & $GL(n,c)$ & ~~~~~$\rightarrow$~~~~~ 
                                                & $GL(n,q)$\\
          & &           & & $\downarrow$  \\
          & &           & & $GL(n,{\cal Q}_c)$\\
          & &           & & $\downarrow$  \\
          & &           & & $GL(n,{\cal Q}_r)$~.
\end{tabular}
\end{center}
Before discussing the groups $GL(n,{\cal Q}_r)$ and $GL(n,{\cal Q}_c)$, we 
introduce a new definition of transpose for quaternionic matrices which will 
allow us to overcome the difficulties due to the noncommutative nature 
of the quaternionic field 
(our definition, applying to 
standard quaternions, will be extended to complex and real linear 
quaternions).

The customary convention of defining the transpose $M^t$ of the matrix $M$ 
is
\[ (M^t)_{rs} = M_{sr} ~.\]
In general, however, for two quaternionic matrices $M$ and $N$ one has
\[ (MN)^t \neq N^t M^t ~,\]
whereas this statement hold as an equality for complex matrices. For example, 
the matrices
\begin{equation}
\left[ \left( \begin{array}{cc} m_1 & m_2\\ m_3 & m_4 \end{array} \right)
\left( \begin{array}{cc} n_1 & n_2\\ n_3 & n_4 \end{array} \right)
\right]^t =
\left( \begin{array}{cc} m_1n_1+m_2n_3~ & ~m_3n_1+m_4n_3 \\ 
                         m_1n_2+m_2n_4~ & ~m_3n_2+m_4n_4 \end{array} \right)
\end{equation}
and
\begin{equation}
\left( \begin{array}{cc} n_1 & n_2\\ n_3 & n_4 \end{array} \right)^t
\left( \begin{array}{cc} m_1 & m_2\\ m_3 & m_4 \end{array} \right)^t =
\left( \begin{array}{cc} n_1m_1+n_3m_2~ & ~n_1m_3+n_3m_4 \\ 
                         n_2m_1+n_4m_2~ & ~n_2m_3+n_4m_4 \end{array} \right)
\end{equation}
are equal only if we use a commutative number field. How can we define 
orthogonal quaternionic groups? 
By looking at the previous example, we see that the 
problem arises in the different position of factors $m_{1,2,3,4}$ and
$n_{1,2,3,4}$ in the elements of our matrices. The solution is very simple once
seen. It is possible to give a quaternionic transpose which reverses the 
order of factors and {\em goes back} to the usual definition for complex 
numbers, 
$c^t=c$. We define
\begin{equation}
\label{tras}
q^t=x_0 + e_1 x_1 -e_2 x_2 + e_3 x_3~.
\end{equation}
In this way, the transpose of a product of two quaternions $q$ and $p$ is the 
product of the transpose quaternions in reverse order
\[ (qp)^t=p^tq^t~.\]
The proof is straightforward if we recognize the following relation 
between transpose 
$q^t$  and conjugate  $q^{\dag}$,
\[ q^t=-e_2q^{\dag}e_2~.\]
Thus, in the quaternionic world the convention of defining the transpose 
$M^t$ of the matrix $M$ will be
\[
(M^t)_{rs} = M^{~~t}_{sr} ~,
\]
or equivalently
\begin{equation}
\label{m2tras}
 M^t = -e_2 M^{\dag} e_2 ~,
\end{equation}
where $M^{\dag}$ is defined in the standard way as
\[
(M^{\dag})_{rs} = M^{~~\dag}_{sr} ~.
\]
With this new definition of quaternionic transpose, the relation
\begin{eqnarray*} 
(MN)^t & = & -e_2 (MN)^{\dag} e_2 = -e_2 N^{\dag} M^{\dag} e_2\\
       & = & (-e_2 N^{\dag} e_2) (-e_2 M^{\dag} e_2)\\ 
       & = & N^{t} M^{t} 
\end{eqnarray*}
also holds for noncommutative numbers.

Noting that under the transpose operation we have $e_{1,3}^t=e_{1,3}$ 
and $e_2^t=-e_2$, we can immediately generalize the definition of transpose 
conjugation to complex and real linear quaternionic operators 
\begin{eqnarray*} 
{\cal Q}_c^t  & =  & q_0^t + q_1^t \mid e_1 ~,\\
{\cal Q}_r^t  & =  & q_0^t + q_1^t \mid e_1 - q_2^t \mid e_2 + 
                             q_3^t \mid e_3 ~.
\end{eqnarray*}
The fundamental property of reversering the order of factors 
for the transpose of quaternionic products
\begin{eqnarray*}  
({\cal Q}_c {\cal P}_c)^t & = & 
[q_0p_0-q_1p_1+(q_0p_1+q_1p_0) \mid e_1]^t \\
                          & = &
p_0^tq_0^t-p_1^tq_1^t+(p_0^tq_1^t+p_1^tq_0^t) \mid e_1\\
                          & = & 
(p_0^t + p_1^t \mid e_1)(q_0^t + q_1^t \mid e_1)\\ 
                          & = & 
{\cal P}_c^t {\cal Q}_c^t  ~,\\ 
({\cal Q}_r {\cal P}_r)^t & = & 
       p_0^tq_0^t - p_1^tq_1^t - p_2^t q_2^t - p_3^tq_3^t +\\
 &   & (p_0^tq_1^t + p_1^tq_0^t + p_2^t q_3^t - p_3^tq_2^t) \mid e_1 -\\
 &   & (p_0^tq_2^t + p_2^tq_0^t + p_3^t q_1^t - p_1^tq_3^t) \mid e_2 +\\
 &   & (p_0^tq_3^t + p_3^tq_0^t + p_1^t q_2^t - p_2^tq_1^t) \mid e_3 \\
 & = & (p_0^t + p_1^t \mid e_1 - p_2^t \mid e_2 + p_3^t \mid e_3) \times \\ 
 &   & (q_0^t + q_1^t \mid e_1 - q_2^t \mid e_2 + q_3^t \mid e_3)\\
 & = & {\cal P}_r^t {\cal Q}_r^t  
\end{eqnarray*}
is again preserved.

In discussing the classification of the classical (matrix) groups, it is 
necessary to introduce one additional concept: the {\em metric}. A metric
function on a vector space is a mapping of a pair of vectors into a number 
field ${\cal F}$ (${\cal F}\equiv {\cal R}/{\cal C}$ for real/complex  
linear operators, see below). Let us now recall the following theorem: 
{\em The subset of transformations of basis in $V_n$ which 
preserves the mathematical structure of a metric forms a subgroup of general 
linear groups}. 
\begin{center}
\begin{tabular}{llll}
                   & ~~~bilinear symmetric      &        
                                             & ~~~{\em orthogonal}\\
Groups preserving  & ~~~bilinear antisymmetric  & ~~~metrics are called 
                                             & ~~~{\em symplectic}\\
                   & ~~~sesquilinear symmetric  &
                                             & ~~~{\em unitary}~. 
\end{tabular}
\end{center}
The previous theorem is valid for all real and complex 
metric-preserving matrix groups. It is also valid for quaternionic groups 
that preserve sesquilinear
metrics, since two quaternions obey $(q_1q_2)^{\dag}=q_2^{\dag}q_1^{\dag}$. It 
is not true for quaternionic matrices and bilinear metrics, since two 
quaternions do not generally commute. Nevertheless, it is still possible to 
associate subgroups of  $GL(n,q)$ with groups that preserve bilinear metrics.
In the literature this is done in the following way. ``{\sl Each quaternion
in $GL(n,q)$ is replaced by the corresponding $2\times 2$ complex matrix 
using the translation rules~(\ref{ident}). 
The subset of matrices in this complex $2n\times 2n$
matrix representation of $GL(n,q)$ that leaves invariant a bilinear metric 
forms a group, since the theorem is valid for bilinear metrics on complex
linear vector spaces. We can associate an $n\times n$ quaternion-valued 
matrix with each $2n\times 2n$ complex-valued matrix in the resulting groups
that preserve bilinear metrics in the space ${\cal C}_{2n}$, which is a 
representation for the space ${\cal H}_n$}'' - {\em Gilmore}~\cite{GIL}. 

Once we write our complex matrix, we can trivially obtain the generators of 
complex orthogonal groups in a standard manner and then we can translate 
back into quaternionic language. But this is {\em surely a laborious 
procedure}. Defining an appropriate transpose for quaternionic 
numbers~(\ref{tras}), we can overcome the just-cited difficulty. Besides, 
using the symplectic representation~(\ref{sym}), the most general 
transformation (on quaternionic states) will be necessarily represented by 
complex linear quaternionic operators, ${\cal Q}_c$, and for the invariant 
metric we have to require a ``complex'' projection
\[ (q^t q)_c = [(c_1 - e_2 c_2^*)(c_1 + e_2 c_2)]_c = c_1^2 + c_2^2~.\] 
We wish to emphasize that the introduction of the imaginary unit $1\mid e_1$
in complex linear quaternionic operators
\[ (1\mid e_1)^{\dag} = -1\mid e_1~,\]
necessarily implies a complex inner product. The ``new'' imaginary units 
$1\mid e_1$ represents an antihermitian operator, and so it must verify
\begin{center}
\begin{tabular}{ccc} 
$\int~(A\psi)^{\dag}\varphi$ & ~~~$=$~~~ &  $-\int~ \psi^{\dag} A \varphi$\\
         $\downarrow$        &           &     $\downarrow$\\
$\int~(\psi e_1)^{\dag}\varphi$ & ~~~$=$~~~ &  
$-\int~ \psi^{\dag} \varphi e_1~.$
\end{tabular}
\end{center}
The previous relation is true only if we adopt a {\em complex projection} 
\[ \int_c \, \equiv \, \frac{1-e_1 \mid e_1}{2} \, \int ~,\]
for
the inner products
\[ \int_c~(\psi e_1)^{\dag}\varphi  =- e_1 \int_c~ \psi^{\dag} \varphi =
- \int_c~ \psi^{\dag} \varphi e_1~.\]

The generators of the unitary and  orthogonal groups 
satisfy the following constraints
\begin{center}
\begin{tabular}{ll}
Groups:~~~~~~~~~~~~~~ & Generators:\\
Unitary               & $A+A^{\dag}=0$ ,\\
Orthogonal            & $A+A^t=0$~.
\end{tabular}
\end{center}
For one-dimensional quaternionic groups, we find  
\begin{center}
\begin{tabular}{|l|l|}
\hline
~~Groups:~~~~~~~~~~~~~~  & ~~Generators:\\
\hline
~~$U(1,q)$               & ~~$e_1~,e_2~,e_3$ ,\\
~~$U(1,{\cal Q}_c)$      & ~~$e_1~,e_2~,e_3~,1\mid e_1 $ ,\\
~~$U(1,{\cal Q}_r)$      & ~~$e_1~,e_2~,e_3~,
                              1\mid e_1~,1\mid e_2~,1\mid e_3$ ,\\
\hline
~~$O(1,q)$               & ~~$e_2$ ,\\
~~$O(1,{\cal Q}_c)$      & ~~$e_2~,e_2\mid e_1 $ ,\\
~~$O(1,{\cal Q}_r)$      & ~~$e_2~,e_1\mid e_2~,e_3\mid e_2~,
                          1\mid e_2~,e_2\mid e_1~,e_2\mid e_3$ .~~\\
\hline
\end{tabular}
\end{center}
At this point, we make a number of observations:

1. - The difference between orthogonal and unitary groups is manifest for 
complex linear quaternionic groups because of the different numbers of 
generators. 

2. - Orthogonal and unitary real linear quaternionic groups have the same 
number of generators.

3. - The groups $U(n_+,n_-,r)$ and $O(n_+,n_-,r)$ are identical
(there is no difference between bilinear and sesquilinear metrics in a real 
vector space) and this suggest a possible link between $U(n,{\cal Q}_r)$ 
and $O(n,{\cal Q}_r)$.

4. - For real linear quaternionic groups, the invariant metric requires a 
``real'' projection (note that $1\mid e_{1,2,3}$ represent antihermitian 
operators only for real inner products).

\noindent Let us show the ``real'' invariant metric for $U(1,{\cal Q}_r)$
and $O(1,{\cal Q}_r)$, 
\begin{eqnarray*} 
(q^{\dag} q)_r &  = &  
[(x_0 -e_1x_1-e_2x_2-e_3x_3)(x_0 +e_1x_1+e_2x_2+e_3x_3)]_r=
x_0^2+x_1^2+x_2^2+x_3^2~,\\
(q^t q)_r &  = &  
[(x_0 +e_1x_1-e_2x_2+e_3x_3)(x_0 +e_1x_1+e_2x_2+e_3x_3)]_r=
x_0^2-x_1^2+x_2^2-x_3^2~.
\end{eqnarray*}
We can immediately recognize the invariant metric of $O(4,r)$ and $O(2,2,r)$.
To complete the analogy between one-dimensional real linear quaternionic 
operators and $4$-dimensional real matrices, we observe that besides the
subgroups related to the ${\dag}$-conjugation (where the sign of all three 
imaginary units is changed)
and the $t$-conjugation (where only $e_2\rightarrow -e_2$), we can define a 
new subgroup which leaves invariant the following real metric
\[ (q^{\dag} g q)_r~,\]
where
\[ g = - \frac{1}{2} \, (1 + e_1 \mid e_1 + e_2 \mid e_2 + e_3 \mid e_3)~.\]
Explicitly,
\[ (q^{\dag} g q)_r = 
[(x_0 -e_1x_1-e_2x_2-e_3x_3)(x_0 -e_1x_1-e_2x_2-e_3x_3)]_r=
x_0^2-x_1^2-x_2^2-x_3^2~.\]
So the new subgroup, $\tilde{O}(1,{\cal Q}_r)$, represents
the one-dimensional quaternionic counterpart of the Lorentz 
group~\cite{SR}. We observe that the $*$-conjugation, obtained by 
changing the sing of two imaginary units, 
\[ q^{*} = x_0 -e_1x_1 - e_2x_2 + e_3x_3~,\]
also gives a real invariant metric related to $O(3,1,r)$
\[ (q^{*} q)_r =   
[(x_0 -e_1x_1-e_2x_2+e_3x_3)(x_0 +e_1x_1+e_2x_2+e_3x_3)]_r=
x_0^2+x_1^2+x_2^2-x_3^2~.\]
Nevertheless, it is not suitable in defining a new subgroup because
\[ (qp)^{*}= q^{*} p^{*}~.\]

The classical groups which occupy a central place in group representation
theory and have many applications in various branches of Mathematics and
Physics are the unitary, special unitary, orthogonal, and symplectic groups.
In order to define special groups, we must define an appropriate
trace for our matrices. In fact, for noncommutative numbers the trace of the
product of two numbers is not the trace of the product with reversed factors.
With complex linear quaternions we have the possibility to give a new 
definition of ``complex'' trace ($Tr$) by
\begin{equation}
\label{ctrace}
Tr~{\cal Q}_c= \mbox{Re} \, q_0 + e_1 \, \mbox{Re} \, q_1~.
\end{equation}
Such a definition implies that for any two complex linear quaternionic 
operators ${\cal Q}_c$ and ${\cal P}_c$
\[ Tr~({\cal Q}_c {\cal P}_c)= Tr~({\cal P}_c {\cal Q}_c)~. \]
For real linear quaternions we need to use the standard definition of ``real''
trace ($tr$)
\begin{equation}
\label{rtrace}
tr~{\cal Q}_r= \mbox{Re} \, q_0~,
\end{equation}
since the previous ``complex'' definition~(\ref{ctrace}) gives
\[ Tr~({\cal Q}_r{\cal P}_r) \neq Tr~({\cal P}_r {\cal Q}_r)~. \]
Explicitly, we find
\begin{eqnarray*}
Tr~({\cal Q}_r{\cal P}_r) & = & 
                          \mbox{Re}~(q_0p_0 - q_1p_1 - q_2 p_2 - q_3p_3) +\\
                           &  & 
                      e_1~\mbox{Re}~(q_0p_1 + q_1p_0 - q_2 p_3 + q_3p_2)~,\\
Tr~({\cal }P_r {\cal Q}_r) & = & 
                          \mbox{Re}~(p_0q_0 - p_1q_1 - p_2 q_2 - p_3q_3) +\\
                           &  & 
                      e_1~\mbox{Re}~(p_0q_1 + p_1q_0 - p_2 q_3 + p_3q_2)~.
\end{eqnarray*}

We recall that the generators of the unitary, special unitary, 
orthogonal groups must satisfy the following conditions~\cite{COR}
\begin{center}
\begin{tabular}{lll}
$U(n)$   & ~~~~~~~$A+A^{\dag}=0$                   ,&                 \\
$SU(n)$  & ~~~~~~~$A+A^{\dag}=0$                   , &~~~ $Tr~A=0$ ,  \\
$O(n)$   & ~~~~~~~$A+A^{t}=0$                      . &                 
\end{tabular}
\end{center}
These conditions also apply for quaternionic groups.

For complex symplectic groups we find
\begin{center}
\begin{tabular}{lll}
$Sp(2n)$ & ~~~~~~~${\cal J}A+A^{t} {\cal J}=0$  , & 
\end{tabular}
\end{center}
where
\[ {\cal J} = \left( \begin{array}{cc} 
\bbox{0}_{n\times n} & \bbox{1}_{n\times n}\\
-\bbox{1}_{n\times n} & \bbox{0}_{n\times n}
              \end{array} \right)~.\]
Working with quaternionic numbers, we can construct a group preserving a 
non-singular antisymmetric metric, for $n$ odd as well 
as $n$ even. Thus for quaternionic symplectic groups we have
\begin{center}
\begin{tabular}{lll}
$Sp(n)$ & ~~~~~~~${\cal J}A+A^{t} {\cal J}=0$  , & 
\end{tabular}
\end{center}
with
\[ {\cal J}_{2n\times 2n} = \left( \begin{array}{cc} 
\bbox{0}_{n\times n} & \bbox{1}_{n\times n}\\
-\bbox{1}_{n\times n} & \bbox{0}_{n\times n}
              \end{array} \right)~, ~~~~~~~ 
{\cal J}_{(2n+1)\times (2n+1)} = \left( \begin{array}{ccc} 
\bbox{0}_{n\times n} & \bbox{0}_{n\times 1}  & \bbox{1}_{n\times n}\\
\bbox{0}_{1\times n} & e_2  & \bbox{1}_{1\times n}\\
-\bbox{1}_{n\times n} & \bbox{0}_{n\times 1} & \bbox{0}_{n\times n}
              \end{array} \right)~.
\]

The generators of one-dimensional groups with complex and real linear
quaternions are
\begin{center}
\begin{tabular}{|lllcl|}
\hline
~~$U(1,{\cal Q}_c)$   & ~~~~~~~${\cal Q}_c+{\cal Q}_c^{\dag}=0$ ,
                    &                 
                    & ~~~$\rightarrow$~~~
                    & ~~~~~~~ $A=e_{1,2,3},~1\mid e_1$ , \\
~~$SU(1,{\cal Q}_c)$  & ~~~~~~~${\cal Q}_c+{\cal Q}^{\dag}=0$ ,
                    & ~~~$Tr~{\cal Q}_c=0$ ,                
                    & ~~~$\rightarrow$~~~
                    & ~~~~~~~ $A=e_{1,2,3}$ , \\
~~$O(1,{\cal Q}_c)$   & ~~~~~~~${\cal Q}_c+{\cal Q}_c^t=0$ ,
                    &                 
                    & ~~~$\rightarrow$~~~
                    & ~~~~~~~ $A=e_2,~e_2\mid e_1$ , \\
~~$Sp(1,{\cal Q}_c)$  
                    & ~~~~~~~$e_2{\cal Q}_c+{\cal Q}_c^t e_2=0$ ,
                    &                 
                    & ~~~$\rightarrow$~~~
                    & ~~~~~~~ $A=e_{1,2,3},
                               ~e_{1,2,3} \mid e_1$ , \\
\hline
~~$U(1,{\cal Q}_r)$   & ~~~~~~~${\cal Q}_r+{\cal Q}_r^{\dag}=0$ ,
                    &                 
                    & ~~~$\rightarrow$~~~
                    & ~~~~~~~ $A=e_{1,2,3},
                               ~1\mid e_{1,2,3}$ , ~~\\
~~$O(1,{\cal Q}_r)$   & ~~~~~~~${\cal Q}_r+{\cal Q}_r^t=0$ ,
                    &                 
                    & ~~~$\rightarrow$~~~
                    & ~~~~~~~ $A=e_2,~e_{1,3}\mid e_2,
                               ~1\mid e_2,~e_2\mid e_{1,3}$ , ~~\\
~~$\tilde{O}(1,{\cal Q}_r)$   & ~~~~~~~$g{\cal Q}_r+{\cal Q}_r^{\dag} g=0$ ,
                    &                 
                    & ~~~$\rightarrow$~~~
                    & ~~~~~~~ $A=e_k-1\mid e_k,~e_i\mid e_j - e_j\mid e_i,$\\
                    & 
                    &                 
                    & 
                    & ~~~~~~~ 
~~~~~~ {\footnotesize $i,j,k=1,2,3$ and $i\neq j$} , ~~\\
~~$Sp(1,{\cal Q}_r)$  & ~~~~~~~$e_2 {\cal Q}_r
                      +{\cal Q}_r^t e_2=0$ , 
                    &                 
                    & ~~~$\rightarrow$~~~
                    & 
                 ~~~~~~~ $A=e_{1,2,3},~1\mid e_2,~e_{1,2,3} \mid e_{1,3}$ .\\
\hline
\end{tabular}
\end{center}

We conclude our classification of quaternionic groups  
giving the general formulas for counting the 
generators of generic $n$-dimensional groups as function of $n$.
\begin{center}
\begin{tabular}{|c|}
\hline
~~Dimensionalities of quaternionic groups~~\\
\hline
\end{tabular}
\end{center}
\begin{center}
\begin{tabular}{|lclcr|}
\hline
~~$U(n,q)$          & ~$\leftrightarrow$~ & $USp(2n,c)$ & ~~~~~~~~~~~ & 
$n(2n+1)$ ,~~\\
~~$U(n,{\cal Q}_c)$ & ~$\leftrightarrow$~ & $U(2n,c)$ & ~~~~~~~~~~~ & 
$4n^2$~ ,~~\\
~~$U(n,{\cal Q}_r)$ & ~$\leftrightarrow$~ & $O(4n,r)$ & ~~~~~~~~~~~ & 
$2n(4n-1)$ ,~~\\
\hline
~~$SU(n,q)$          & ~$\equiv$~ & $U(n,q)$ & ~~~~~~~~~~~ & 
 \\
~~$SU(n,{\cal Q}_c)$ & ~$\leftrightarrow$~ & $SU(2n,c)$ & ~~~~~~~~~~~ & 
$4n^2-1$~ ,~~\\
~~$SU(n,{\cal Q}_r)$ & ~$\equiv$~ & $U(n,{\cal Q}_r)$ & ~~~~~~~~~~~ & 
\\
\hline
~~$O(n,q)$          & ~$\leftrightarrow$~ & $SO^{*}(2n,c)$ & ~~~~~~~~~~~ & 
$n(2n-1)$ ,~~\\
~~$O(n,{\cal Q}_c)$ & ~$\leftrightarrow$~ & $O(2n,c)$ & ~~~~~~~~~~~ & 
$2n(2n-1)$ ,~~\\
~~$O(n,{\cal Q}_r)$ & ~$\leftrightarrow$~ & $O(2n_+,2n_-,r)$ & ~~~~~~~~~~~ & 
$2n(4n-1)$ ,~~\\
~~$\tilde{O}(n,{\cal Q}_r)$ 
                  & ~$\leftrightarrow$~ & $O(3n_+,n_-,r)$ & ~~~~~~~~~~~ & 
$2n(4n-1)$ ,~~\\
\hline
~~$Sp(n,q)$          & ~$\leftrightarrow$~ & $USp(2n,c)$ & ~~~~~~~~~~~ & 
$n(2n+1)$ ,~~\\
~~$Sp(n,{\cal Q}_c)$ & ~$\leftrightarrow$~ & $Sp(2n,c)$ & ~~~~~~~~~~~ & 
$2n(2n+1)$ ,~~\\
~~$Sp(n,{\cal Q}_r)$ & ~$\leftrightarrow$~ & $Sp(4n,r)$ & ~~~~~~~~~~~ & 
$2n(4n+1)$ .~~\\
\hline
\end{tabular}
\end{center}


\section{Octonionic Groups}
\label{s5}

The introduction of real linear quaternionic barred operators, with their $16$
real parameters, naturally suggests a link between ${\cal Q}_r$ and a generic
$4\times 4$ real matrix. Besides, complex linear quaternionic operators, 
${\cal Q}_c$, are characterized by four complex parameters, the same number of
parameters which characterizes two-dimensional complex matrices.

If we try to repeat such a counting for real linear octonionic barred 
operators, due to the nonassociativity, we find 106 real parameters. 
Nevertheless, as 
remarked in section~\ref{s3}, we can express right-barred operators by 
left-barred operators reducing to 64 the previous counting. So, we don't have
apparently  any problem in repeating the considerations of  
section~\ref{s4}. Yet, for complex linear octonionic operators, this will not 
be so easy (the nonassociativity of the 
octonionic field gives some drawbacks). 

Let us try to generalize the 
symplectic complex representation~(\ref{sym}) of a quaternionic state to an 
octonionic state
\[ o = c_1 + e_2 c _2 + e_4 c_3 + e_6 c_4 ~~~~~~~~~ (c_{1,2,3,4} \in {\cal C})
   ~, \]
by the complex matrix
\begin{equation}
\label{sym2}
o ~\leftrightarrow ~ \left( \begin{array}{c} c_1 \\ c_2 \\ c_3 \\ c_4
                            \end{array}  \right)~.
\end{equation}
We immediately see that the barred quaternionic imaginary unit $1\mid e_1$,
when applied to an octonionic state, gives
\[ (1\mid e_1) \, o = e_1 c_1 + e_2 (e_1c _2) + e_4 (e_1 c_3) + 
                      e_6 (e_1 c_4) ~. \]
We don't have problems with the nonassociativity of octonions because 
$e_1e_2e_3$, $e_1e_4e_5$, $e_1e_7e_6$ form quaternionic subalgebras and 
so the associativity is restored. 
Thus, the barred imaginary unit $1\mid e_1$ can be identified with the 
4-dimensional complex matrix $i\openone_{4\times 4}$. The only difference 
between quaternions and octonions is thus represented (as expected) by the 
dimension of the complex matrix. In order to complete the translation we need
to find the octonionic counterpart of~(\ref{ident}). It is here that we have 
problems. Let us observe the following embarrassing situation. The action of 
$e_2$ on an octonionic state is
\begin{eqnarray*} 
e_2 o  & = & e_2 c_1 + e_2 (e_2 c _2) + e_2 (e_4 c_3) + e_2 (e_6 c_4)\\
       & = & -c_2 + e_2 c_1 - e_4 c_4^* + e_6 c_3^* ~. 
\end{eqnarray*}
So we should have 
\[ e_2 o ~\leftrightarrow ~ M_{(e_2)}  
                    \left( \begin{array}{c} c_1 \\ c_2 \\ c_3 \\ c_4
                           \end{array}  \right)
                           \stackrel{?}{=}
                    \left( \begin{array}{c} $-$c_2 \\ c_1 \\ $-$c_4^* \\ c_3^*
                           \end{array}  \right)~.
\]
Obviously we cannot find a complex (left-acting) matrix $M_{(e_2)}$ which 
corresponds to the 
octonionic imaginary unit $e_2$. We also note that the 
octonionic imaginary unit $e_2$ does not represent an antihermitian operator, 
as desired to generalize the quaternionic translation rules~(\ref{ident}). 
In fact, for nonassociative numbers, we have
\begin{center}
\begin{tabular}{ccc} 
$\int_c~(A\psi)^{\dag}\varphi$ & ~~~~~~ &  $-\int_c~ \psi^{\dag} (A 
\varphi)$\\
         $\downarrow$        &           &     \\
$\int_c~(e_2 \psi )^{\dag}\varphi$ &  &  $\downarrow$ \\
         $\downarrow$        &           &     \\
$-\int_c~ (\psi^{\dag} e_2) \varphi$ & ~~~$\stackrel{?}{=}$~~~ &
$-\int_c~ \psi^{\dag} (e_2 \varphi )$ .
\end{tabular}
\end{center}
For complex linear octonionic operators $e_1$ and $1\mid e_1$ represent 
antihermitian imaginary units, whereas the remaining $e_{2,...,7}$ are not 
antihermitian operators. The proof is quoted in appendix~C. 

We wish to obtain the octonionic generalization of ``$e_2$'', which 
translates to
\begin{equation}
\label{e2}
e_2 ~ \leftrightarrow ~ \left( \begin{array}{cc} 
0 & $-$1 \\ 
1 & 0 
\end{array} \right)~,
\end{equation}
when considered a quaternionic number and which obviously represents an 
antihermitian 
operator in the octonionic world. The solution is
\begin{equation}
\label{e22}
\mbox{``}e_2\mbox{''} ~ \equiv ~  
e_2 + \frac{e_3~)~e_1 - e_3~(~e_1}{2}
~ \leftrightarrow ~ 
\left( \begin{array}{cccc} 
0 & $-$1 & 0 & 0 \\ 
1 & 0    & 0 & 0 \\
0 & 0    & 0 & 0 \\
0 & 0    & 0 & 0                              
\end{array} \right)~.
\end{equation}
Eq.~(\ref{e22}) represents the octonionic counterpart of Eq.~(\ref{e2}), and
goes back to the usual definition for quaternionic numbers, in fact
\[ \frac{e_3~)~e_1 - e_3~(~e_1}{2} \, q \equiv 0 ~. 
\]
In a similar manner we can construct the octonionic counterpart of the 
quaternionic hermitian operator
\[ e_3\mid e_1 ~ \leftrightarrow ~ \left( \begin{array}{cc} 
0 & 1 \\ 
1 & 0 
\end{array} \right)~.
\]
Explicitly,
\[ \frac{e_3~)~e_1 + e_3~(~e_1}{2}
~ \leftrightarrow ~ 
\left( \begin{array}{cccc} 
0 & 1    & 0 & 0 \\ 
1 & 0    & 0 & 0 \\
0 & 0    & 0 & 0 \\
0 & 0    & 0 & 0                              
\end{array} \right)~.
\]

In a previous paper~\cite{OREP} we discussed the link between $GL(4,c)$ and
complex linear octonionic operators. Nevertheless, in that paper we were not
able to connect directly the 16 (complex) basis elements of $GL(4,c)$ with
complex linear octonionic operators. ``{\sl The 32 (real) basis 
elements of $GL(4,c)$ can be extracted from the 64 generators of 
$GL(8,r)$}''~\cite{OREP}. We can now directly connect the antihermitian 
operators
\[ 
\left( \begin{array}{cccc} 
0 & $-$1    & 0 & 0 \\ 
1 & 0    & 0 & 0 \\
0 & 0    & 0 & 0 \\
0 & 0    & 0 & 0                              
\end{array} \right)~~~~~,~~~~~
\left( \begin{array}{cccc} 
0 & 0    & $-$1 & 0 \\ 
0 & 0    & 0 & 0 \\
1 & 0    & 0 & 0 \\
0 & 0    & 0 & 0                              
\end{array} \right)~~~~~,~~~~~
\left( \begin{array}{cccc} 
0 & 0    & 0 & $-$1 \\ 
0 & 0    & 0 & 0 \\
0 & 0    & 0 & 0 \\
1 & 0    & 0 & 0                              
\end{array} \right)~~~~~,~~~~~...
\]
with the complex linear octonionic operators
\[
e_2 + \frac{e_3~)~e_1 - e_3~(~e_1}{2} ~~~,~~~
e_4 + \frac{e_5~)~e_1 - e_5~(~e_1}{2} ~~~,~~~
e_6 - \frac{e_7~)~e_1 - e_7~(~e_1}{2} ~~~,~~~...
\]
and the hermitian operators
\[
\left( \begin{array}{cccc} 
0 & 1    & 0 & 0 \\ 
1 & 0    & 0 & 0 \\
0 & 0    & 0 & 0 \\
0 & 0    & 0 & 0                              
\end{array} \right)~~~~~,~~~~~
\left( \begin{array}{cccc} 
0 & 0    & 1 & 0 \\ 
0 & 0    & 0 & 0 \\
1 & 0    & 0 & 0 \\
0 & 0    & 0 & 0                              
\end{array} \right)~~~~~,~~~~~
\left( \begin{array}{cccc} 
0 & 0    & 0 & $-$1 \\ 
0 & 0    & 0 & 0 \\
0 & 0    & 0 & 0 \\
$-$1 & 0    & 0 & 0                              
\end{array} \right)~~~~~,~~~~~...
\]
with
\[
\frac{e_3~)~e_1 + e_3~(~e_1}{2} ~~~,~~~
\frac{e_5~)~e_1 + e_5~(~e_1}{2} ~~~,~~~
\frac{e_7~)~e_1 + e_7~(~e_1}{2} ~~~,~~~...
\]
In this way we can immediately recognize the antihermitian operators as 
generators of $U(1,{\cal O}_c)$, and so we obtain the octonionic 
(one-dimensional) counterpart of $U(4,c)$.

We conclude this section with some considerations concerning the 
octonionic groups.

1. - The nonassociativity is not a problem for real/complex octonionic 
operators because we have a correspondence between such  octonionic 
operators and (associative) general linear groups, $GL(8,r)$/$GL(4,c)$.

2. - By our translation rules, we can immediately obtain the octonionic 
counterpart of $GL(8n,r)$ and $GL(4n,c)$. Thus, we can translated a part of 
the standard real/complex groups.

3. - We must admit some technical difficulties in manipulating the octonionic 
field, see for example the nonantihermiticity (for complex linear operators)
of the ``standard'' octonionic imaginary units $e_{2,...,7}$.

4. - The considerations of section~\ref{s4} (transpose, trace, ...) can be 
extended to octonionic fields. For complex linear octonionic operators, we
must work with the new octonionic imaginary units ``$e_2$'', ``$e_4$'', 
``$e_6$'', etc.


\section{Conclusions}
\label{s6}

The more exciting possibility that quaternionic or octonionic equations will
eventually play a significant role in Mathematics and Physics is synonymous, 
for some physicist, with
the advent of a revolution in Physics comparable to that of Quantum Mechanics.

For example, Adler suggested~\cite{ADLP} that the color degree of freedom
postulated in the Harari-Shupe model~\cite{H,S} (where we can think of
quarks and leptons as composites of other more fundamental fermions, preons) 
could be sought in a noncommutative extension of the complex field.
Surely a stimulating idea. Nevertheless, we think that it would be very 
strange if standard Quantum 
Mechanics did not permit a quaternionic or octonionic description other than
in the trivial sense that complex numbers are contained within the quaternions
or octonions. 

In the last few years much progress has been achieved in manipulating
such fields. We quote the quaternionic version of electroweak 
theory~\cite{QET}, where the Glashow group is expressed by the one-dimensional
quaternionic group $U(1,q)\mid U(1,c)$, quaternionic GUTs~\cite{GUT} and
Special Relativity, where the Lorentz group is represented by 
$\tilde{O}(1,{\cal Q}_r)$.  We also recall new possibilities related to the 
use of octonions in Quantum Mechanics~\cite{OQM}, in particular in writing a 
one-dimensional octonionic Dirac equation~\cite{ODE}. The link between 
octonionic and quaternionic versions of standard Quantum Physics is represented
by the use of a complex geometry~\cite{REM}.

In this paper we observe that beyond the study of matrix groups with 
``simple'' quaternionic elements, $q$, one can consider more general groups 
with matrix elements of the form ${\cal Q}_{r/c}$. To the best of our 
knowledge these more general matrix groups have not been studied in the 
literature.  We overcome the problems arising in the definitions of transpose, 
determinant and trace for quaternionic matrices. For octonionic fields we must
admit a more complicated situation, yet our discussion can be also proposed  
for nonassociative numbers. 

Finally, we hope that this paper emphasizes the possibility of using 
hypercomplex numbers in Mathematics and Physics and it represents an 
important step towards a complete discussion of Hypercomplex Matrix Algebras.

\acknowledgements

The author's work was financially supported by a fellowship of Italian  
CNR and in part by a research grant of Brazilian FINEP. It is a pleasure 
to gratefully knowledge Dr.~J.~E.~Maiorino, Prof.~E.~Capellas de Oliveira,  
and Prof.~L.~Venditte for their warm 
hospitality during the stay at the Instituto de Matem\'atica, Estat\'{\i}stica 
e Computa\c{c}\~ao Cient\'{\i}fica (UNICAMP) where this paper was prepared. 
The author is deeply indebted to director of IMECC, 
Prof.~Waldyr A.~Rodrigues, Jr. for many 
stimulating conversations on Geometric Algebra, and Prof.~P.~Rotelli for  
useful suggestions 
and a critical reading of the manuscript. Finally, the author would like also 
thanks his wife {\em Rita} for her patience and invaluable encouragement.



\section*{Appendix A
\\ Quaternionic Barred Operators and $\bbox{4\times 4}$ Real 
Matrices}

We give the translation rules between quaternionic  
barred operators and $4\times 4$ real matrices:
\begin{eqnarray*}
\bbox{e_1} = \left( \begin{array}{cccc} 
0 & $-$1    & 0 & 0 \\ 
1 & 0    & 0 & 0 \\
0 & 0    & 0 & $-$1 \\
0 & 0    & 1 & 0                              
\end{array} \right)~, &~~~~~~~&
\bbox{1\mid e_1} = \left( \begin{array}{cccc} 
0 & $-$1    & 0 & 0 \\ 
1 & 0    & 0 & 0 \\
0 & 0    & 0 & 1 \\
0 & 0    & $-$1 & 0 
\end{array} \right)~,\\
\bbox{e_2} = \left( \begin{array}{cccc} 
0 & 0    &$-$1 & 0 \\ 
0 & 0    & 0 & 1 \\
1 & 0    & 0 & 0 \\
0 & $-$1    & 0 & 0                              
\end{array} \right)~, &~~~~~~~&
\bbox{1\mid e_2} = \left( \begin{array}{cccc} 
0 & 0    & $-$1 & 0 \\ 
0 & 0    & 0 & $-$1 \\
1 & 0    & 0 & 0 \\
0 & 1    & 0 & 0 
\end{array} \right)~,\\
\bbox{e_3} = \left( \begin{array}{cccc} 
0 & 0    & 0 & $-$1 \\ 
0 & 0    & $-$1 & 0 \\
0 & 1    & 0 & 0 \\
1 & 0    & 0 & 0                              
\end{array} \right)~, &~~~~~~~&
\bbox{1\mid e_3} = \left( \begin{array}{cccc} 
0 & 0    & 0 & $-$1 \\ 
0 & 0    & 1 & 0 \\
0 & $-$1    & 0 & 0 \\
1 & 0    & 0 & 0 
\end{array} \right)~.
\end{eqnarray*}

The remaining rules can be easily constructed noting that
\begin{eqnarray*}
\bbox{e_m}             & ~~\leftrightarrow ~~& L_m ~,\\
\bbox{1 \mid e_m}      & ~~\leftrightarrow ~~& R_m  ~,\\
\bbox{e_m \mid e_n}    & ~~\leftrightarrow ~~& M^{L}_{mn} 
~\equiv ~ R_n L_m  ~,\\
                &      & M^{R}_{mn}  ~\equiv ~ L_m R_n ~.
\end{eqnarray*}


\section*{Appendix B
\\ Octonionic Left-Right  Barred Operators and $\bbox{8\times 8}$ Real 
Matrices}

In this appendix we give the translation rules between octonionic 
left-right  barred operators and $8\times 8$ real matrices. In order to 
simplify our presentation we introduce the following notation:
\begin{mathletters}
\begin{eqnarray}
\{~a, \; b, \; c, \; d~\}_{(1)} ~ \equiv ~
\left( \begin{array}{cccc} a & 0 & 0 & 0\\ 0 & b & 0 & 0\\
0 & 0 & c & 0\\ 0 & 0 & 0 & d \end{array} \right)   & \quad , \quad &
\{~a, \; b, \; c, \; d~\}_{(2)} ~ \equiv ~
\left( \begin{array}{cccc} 0 & a & 0 & 0\\ b & 0 & 0 & 0\\
0 & 0 & 0 & c\\ 0 & 0 & d & 0 \end{array} \right) \quad ,\\
\{~a, \; b, \; c, \; d~\}_{(3)} ~ \equiv ~
\left( \begin{array}{cccc} 0 & 0 & a & 0\\ 0 & 0 & 0 & b\\
c & 0 & 0 & 0\\ 0 & d & 0 & 0 \end{array} \right)   & \quad , \quad &
\{~a, \; b, \; c, \; d~\}_{(4)} ~ \equiv ~
\left( \begin{array}{cccc} 0 & 0 & 0 & a\\ 0 & 0 & b & 0\\
0 & c & 0 & 0\\ d & 0 & 0 & 0 \end{array} \right) \quad ,
\end{eqnarray}
\end{mathletters}
where $a, \; b, \; c, \; d$ and $0$ represent $2\times 2$ real matrices.

From now on, with $\sigma_{1}$, $\sigma_{2}$, $\sigma_{3}$ we represent the 
standard Pauli matrices:
\begin{equation}
\sigma_{1} = \left( \begin{array}{cc} 0 & 1 \\ 1 & 0 \end{array} \right)
 \quad , \quad  
\sigma_{2} = \left( \begin{array}{cc} 0 & -i \\ i & 0 \end{array} \right) 
\quad , \quad  
\sigma_{3} = \left( \begin{array}{cc} 1 & 0 \\ 0 & -1 \end{array} \right) 
\quad .  
\end{equation}

The only necessary translation rules that we need to know 
explicitly are the following
\begin{center}
\begin{tabular}{rcrrrrcrcrrrrc} 
 & & & & & & & & & & & & & \\ 
$\bbox{e_{1}}$ & $~\leftrightarrow~\{$ & $-i\sigma_{2}$, & ~$-i\sigma_{2}$, & 
~$-i\sigma_{2}$, & 
~$i\sigma_{2} ~\}_{(1)}$
&~~~,~~~~& 
$\bbox{1\mid e_{1}}$ & $~\leftrightarrow~\{$ & 
$-i\sigma_{2}$, & ~$i\sigma_{2}$, & 
~$i\sigma_{2}$, & 
~$-i\sigma_{2} ~\}_{(1)}$
&~~~,~~~~\\
$\bbox{e_{2}}$ & $~\leftrightarrow~\{$ & 
$ -\sigma_{3}$, & ~$\sigma_{3}$, & ~$-\openone$, & 
~$\openone ~\}_{(2)}$
&~~~,~~~~& 
$\bbox{1\mid e_{2}}$ & $~\leftrightarrow~\{$ & 
$ -\openone$, & ~$\openone$, & ~$\openone$, & 
~$-\openone ~\}_{(2)}$
&~~~,~~~~\\
$\bbox{e_{3}}$ & $~\leftrightarrow~\{$ & $ -\sigma_{1}$, & ~$\sigma_{1}$, & 
~$-i\sigma_{2}$, & 
~$-i\sigma_{2} ~\}_{(2)}$
&~~~,~~~~& 
$\bbox{1\mid e_{3}}$ & $~\leftrightarrow~\{$ & 
$ -i\sigma_{2}$, & ~$-i\sigma_{2}$, & 
~$i\sigma_{2}$, & 
~$i\sigma_{2} ~\}_{(2)}$
&~~~,~~~~\\
$\bbox{e_{4}}$ & $~\leftrightarrow~\{$ & 
$ -\sigma_{3}$, & ~$\openone$, & ~$\sigma_{3}$, & 
~$-\openone ~\}_{(3)}$
&~~~,~~~~& 
$\bbox{1\mid e_{4}}$ & $~\leftrightarrow~\{$ & 
$ -\openone$, & ~$-\openone$, & ~$\openone$, 
& ~$\openone ~\}_{(3)}$
&~~~,~~~~\\
$\bbox{e_{5}}$ & $~\leftrightarrow~\{$ & 
$ -\sigma_{1}$, & ~$i\sigma_{2}$, & ~$\sigma_{1}$, 
& ~$i\sigma_{2} ~\}_{(3)}$
&~~~,~~~~& 
$\bbox{1\mid e_{5}}$ & $~\leftrightarrow~\{$ & 
$ -i\sigma_{2}$, & ~$-i\sigma_{2}$, & 
~$-i\sigma_{2}$, & 
~$-i\sigma_{2} ~\}_{(3)}$
&~~~,~~~~\\
$\bbox{e_{6}}$ & $~\leftrightarrow~\{$ & 
$ -\openone$, & ~$-\sigma_{3}$, & ~$\sigma_{3}$, & 
~$\openone ~\}_{(4)}$
&~~~,~~~~& 
$\bbox{1\mid e_{6}}$ & $~\leftrightarrow~\{$ & 
$ -\sigma_{3}$, & ~$\sigma_{3}$, & 
~$-\sigma_{3}$, & 
~$\sigma_{3} ~\}_{(4)}$
&~~~,~~~~\\
$\bbox{e_{7}}$ & $~\leftrightarrow~\{$ & 
$ -i\sigma_{2}$, & ~$-\sigma_{1}$, & 
~$\sigma_{1}$, & 
~$-i\sigma_{2} ~\}_{(4)}$
&~~~,~~~~& 
$\bbox{1\mid e_{7}}$ & $~\leftrightarrow~\{$ & $ -\sigma_{1}$, & ~$\sigma_{1}$, & 
~$-\sigma_{1}$, & 
~$\sigma_{1} ~\}_{(4)}$
&~~~.~~~~\\
\end{tabular}
\end{center}

The remaining rules can be easily constructed remembering that
\begin{eqnarray*}
\bbox{e_m}             & ~~\leftrightarrow ~~& L_m ~,\\
\bbox{1 \mid e_m}      & ~~\leftrightarrow ~~& R_m  ~,\\
\bbox{e_m \mid e_m}    & ~~\leftrightarrow ~~& M^{L}_{mm} 
~\equiv ~ R_m L_m  ~,\\
                &      & M^{R}_{mm}  ~\equiv ~ L_m R_m ~,\\  
\bbox{e_m ~)~ e_n}     & ~~\leftrightarrow ~~& M^{L}_{mn} ~\equiv ~  
R_n L_m  ~,\\ 
\bbox{e_m ~( ~e_n}     & ~~\leftrightarrow ~~ & M^{R}_{mn}  ~\equiv ~ 
L_m R_n  ~. 
\end{eqnarray*}
Following this procedure any matrix representation of
right/left barred operators can be obtained. Using 
Mathematica~\cite{MAT},  we proved the linear independence of
the 64 elements which represent the most general octonionic operator  
\[
o_{0}+\sum_{m=1}^{7} o_{m}~)~e_{m} ~ .
\]
Our left  barred operators form a complete basis for any $8\times 8$ real 
matrix and this establishes the isomorphism between $GL(8)$ and left barred 
octonions.


\section*{Appendix C
\\ Antihermiticity properties of complex linear octonionic operators}

Let us consider the action of barred operators on octonionic functions
\[ \psi = c_{1} +e_{2}c_{2}+e_{4}c_{3}+e_{6}c_{4} 
~~~~~~~(c_{1, ...,4} \in {\cal C}) ~ .
\]
With the notation
\[
e_{2} ~\rightarrow ~ \{ -c_{2}, \; c_{1}, \;
-c_{4}^{*}, \; c_{3}^{*} \} ~ ,
\]
we will indicate
\[ 
e_{2}\psi~=~ -c_{2}+e_{2}c_{1}-e_{4}c_{4}^{*}+e_{6}c_{3}^{*} ~ .
\]

The action of a generic octonionic barred operators can be expressed by 
suitable combinations of he action of barred operators $\bbox{e_{m}}$ and 
$\bbox{1\mid e_{m}}$
\begin{center}
\begin{tabular}{rcrrrrcrcrrrrc}
$\bbox{e_{1}}$ & $~\rightarrow~\{$ & $e_{1}c_{1}$, & ~$-e_{1}c_{2}$, & 
~$-e_{1}c_{3}$, & 
~$-e_{1}c_{4} ~\}$
&~~,~~~& 
$\bbox{1\mid e_{1}}$ & 
$~\rightarrow~\{$ & $e_{1}c_{1}$, & ~$e_{1}c_{2}$, & 
~$e_{1}c_{3}$, & 
~$e_{1}c_{4} ~\}$
&~~,~~~\\
$\bbox{e_{2}}$ & $~\rightarrow~\{$ & $ -c_{2}$, & ~$c_{1}$, 
& ~$-c_{4}^{*}$, & ~$c_{3}^{*} ~\}$
&~~,~~~& 
$\bbox{1\mid e_{2}}$ & 
$~\rightarrow~\{$ & $ -c_{2}^{*}$, & ~$c_{1}^{*}$, 
& ~$c_{4}^{*}$, & 
~$-c_{3}^{*} ~\}$
&~~,~~~\\
$\bbox{e_{3}}$ & $~\rightarrow~\{$ & $ -e_{1}c_{2}$, & ~$-e_{1}c_{1}$, 
& ~$-e_{1}c_{4}^{*}$, & 
~$e_{1}c_{3}^{*} ~\}$
&~~,~~~& 
$\bbox{1\mid e_{3}}$ & 
$~\rightarrow~\{$ & $ e_{1}c_{2}^{*}$, & ~$-e_{1}c_{1}^{*}$, 
& ~$e_{1}c_{4}^{*}$, & 
~$-e_{1}c_{3}^{*} ~\}$
&~~,~~~\\
$\bbox{e_{4}}$ & 
$~\rightarrow~\{$ & $ -c_{3}$, & ~$c_{4}^{*}$, 
& ~$c_{1}$, & ~$-c_{2}^{*} ~\}$
&~~,~~~& 
$\bbox{1\mid e_{4}}$ & 
$~\rightarrow~\{$ & $ -c_{3}^{*}$, & ~$-c_{4}^{*}$, 
& ~$c_{1}^{*}$, & 
~$c_{2}^{*} ~\}$
&~~,~~~\\
$\bbox{e_{5}}$ & 
$~\rightarrow~\{$ & $ -e_{1}c_{3}$, & ~$e_{1}c_{4}^{*}$, 
& ~$-e_{1}c_{1}$, & 
~$-e_{1}c_{2}^{*} ~\}$
&~~,~~~& 
$\bbox{1\mid e_{5}}$ & $~\rightarrow~\{$ & $ e_{1}c_{3}^{*}$, & 
~$-e_{1}c_{4}^{*}$, 
&~$-e_{1}c_{1}^{*}$, & 
~$e_{1}c_{2}^{*} ~\}$
&~~,~~~\\
$\bbox{e_{6}}$ & $~\rightarrow~\{$ & $ -c_{4}$, & ~$-c_{3}^{*}$, &
 ~$c_{2}^{*}$, & 
~$c_{1} ~\}$
&~~,~~~&
$\bbox{1\mid e_{6}}$ & 
$~\rightarrow~\{$ & $ -c_{4}^{*}$, & ~$c_{3}^{*}$, 
& ~$-c_{2}^{*}$, & 
~$c_{1}^{*} ~\}$
&~~,~~~\\
$\bbox{e_{7}}$ & 
$~\rightarrow~\{$ & $ e_{1}c_{4}$, & ~$e_{1}c_{3}^{*}$, 
& ~$-e_{1}c_{2}^{*}$, & ~$e_{1}c_{1} ~\}$
&~~,~~~& 
$\bbox{1\mid e_{7}}$ & 
$~\rightarrow~\{$ & $ -e_{1}c_{4}^{*}$, 
& ~$-e_{1}c_{3}^{*}$, 
& ~$e_{1}c_{2}^{*}$, & 
~$e_{1}c_{1}^{*} ~\}$
&~~~.~~~ 
\end{tabular}
\end{center}

We can immediately verify that $e_{1}$ represents an antihermitian operator. 
The antihermiticity of $e_{1}$ is shown if
\begin{equation}
\int_{c} ~ \psi^{\dag} (e_{1}\varphi) = 
-\int_{c} ~ (e_{1}\psi)^{\dag} \varphi ~ .
\end{equation}
In the previous equation the only nonvanishing terms are represented by 
{\em diagonal} terms ($\sim c_{1}^{\dag}z_{1}, 
\; c_{2}^{\dag}z_{2}, \; c_{3}^{\dag}z_{3}, 
\; c_{4}^{\dag}z_{4}$). In fact, {\em off-diagonal} terms, like 
$c_{2}^{\dag}z_{3}, \; c_{3}^{\dag}z_{4}$, are killed 
by the complex projection,
\begin{eqnarray*}
~(c_{2}^{\dag} e_{2})[e_{1}(e_{4}z_{3})] & ~\sim ~ & 
(\alpha_{2}e_{2}+\alpha_{3}e_{3})(\alpha_{4}e_{4}+\alpha_{5}e_{5}) \sim  
\alpha_{6}e_{6} + \alpha_{7} e_{7} ~ , \\ 
~[(c_{3}^{\dag} e_{4}) e_{1}](e_{6}z_{4}) & ~\sim ~& 
~(\beta_{4}e_{4}+\beta_{5}e_{5})(\beta_{6}e_{6}+\beta_{7}e_{7}) \sim  
\beta_{2}e_{2} + \beta_{3} e_{3} ~ , \\
 & & ~~~~~~~~~~~[~\alpha_{2, ..., 7}~\mbox{and}~\beta_{2, ..., 7} \in  
     {\cal R}~] ~ .
\end{eqnarray*}
The diagonal terms give
\begin{eqnarray*}
\int_{c} ~ \psi^{\dag} (e_{1}\varphi) & = &
c_{1}^{\dag}[e_{1}z_{1}]
-(c_{2}^{\dag}e_{2})[e_{1}(e_{2}z_{2})]
-(c_{3}^{\dag}e_{4})[e_{1}(e_{4}z_{3})]
-(c_{4}^{\dag}e_{6})[e_{1}(e_{6}z_{4})] ~ ,\\
-\int_{c} ~ (e_{1}\psi)^{\dag} \varphi  & = &
[c_{1}^{\dag}e_{1}]z_{1}
-[(c_{2}^{\dag}e_{2})e_{1}](e_{2}z_{2})
-[(c_{3}^{\dag}e_{4})e_{1}](e_{4}z_{3})
-[(c_{4}^{\dag}e_{6})e_{1}](e_{6}z_{4}) ~ .
\end{eqnarray*}
The parenthesis in the previous equations are not of relevance since
\begin{center}
\begin{tabular}{lll}
$c_{1}^{\dag}e_{1}z_{1}$ & 
{\footnotesize ~~~$(1, \; e_{1})$} & ~~~is a complex 
number , \\ 
$c_{2}^{\dag} e_{2} e_{1} e_{2}z_{2}$ & 
{\footnotesize ~~~(subalgebra 123)} , & \\ 
$c_{3}^{\dag} e_{4} e_{1} e_{4}z_{3}$ & 
{\footnotesize ~~~(subalgebra 145)} , & \\
$c_{4}^{\dag} e_{6} e_{1} e_{6}z_{4}$ & 
{\footnotesize ~~~(subalgebra 176)}   & 
are quaternionic numbers .
\end{tabular}
\end{center}
The antihermiticity proof of $1\mid e_{1}$ is 
very similar to that of $e_1$. The above-mentioned demonstration does not 
work for the imaginary units 
$e_{2}$, ... , $e_{7}$ (breaking the symmetry between the seven 
octonionic imaginary units). In fact for $e_2$ we find
\begin{eqnarray*}
\int_{c} ~ \psi^{\dag} (e_{2}\varphi) & = &
[(c_{1}^{\dag} - c_{2}^{\dag}e_2 -c_{3}^{\dag}e_4 -c_{4}^{\dag}e_6) 
(-z_2+ e_2 z_1  - e_4 z_4^* + e_6 z_3^*)]_c\\
                                      & = &
-c_{1}^{\dag}z_2 + c_{2}^{\dag}z_1 +   c_{3}^{\dag}z_4^* +  c_{4}^{\dag}z_3^*
~ ,\\
-\int_{c} ~ (e_{2}\psi)^{\dag} \varphi  & = &
[(c_{2}^{\dag} + c_{1}^{\dag}e_2  -c_{4}^{\dag}e_4+c_{3}^{\dag}e_6) 
(z_1+ e_2 z_2  + e_4 z_3 + e_6 z_4)]_c\\
                                      & = &
c_{2}^{\dag}z_1 - c_{1}^{\dag}z_2 +   c_{4}^{\dag}z_3 -  c_{3}^{\dag}z_4~ .
\end{eqnarray*}
A similar proof works for $e_{3,...,7}$.


\end{document}